\documentclass[floatfix,showpacs,preprintnumbers,amsmath,amssymb,aps,twocolumn,superscriptaddress,pra,10pt]{revtex4-1}

\usepackage{amsfonts} \usepackage[pdftex]{graphicx} \usepackage{color}
\usepackage{bm} \usepackage{multirow} 
\unitlength=1mm \newcommand{\figref}[1]{Fig.~\ref{#1}}
\newcommand{\tabref}[1]{Table~\ref{#1}} 
 \newcommand{\diffd}{\text{d}}
 \renewcommand{\vec}[1]{\mathbf{#1}}

 \newcommand{\vac}{|0\rangle}
 
\newcommand{\ket}[1]{|{#1}\rangle} \newcommand{\ud}[1]{c^{\dagger}_{#1\uparrow}}
\newcommand{\dd}[1]{c^{\dagger}_{#1\downarrow}}
\newcommand{\cu}[1]{c_{#1\uparrow}} \newcommand{\cd}[1]{c_{#1\downarrow}}

\begin{document}

\title{Exploring exchange mechanisms with a cold atom gas}
\author{P.O.~Bugnion} \affiliation{Cavendish Laboratory, J.J. Thomson Avenue,
Cambridge, CB3 0HE, United Kingdom} \author{G.J.~Conduit}
\affiliation{Cavendish Laboratory, J.J. Thomson Avenue, Cambridge, CB3 0HE,
United Kingdom} \date{\today}

\begin{abstract} Fermionic atoms trapped in a double well potential are an
  ideal setting to study fundamental exchange mechanisms. We use exact
  diagonalization and complementary analytic calculations to demonstrate
  that two trapped fermions deliver a minimal model of the direct exchange
  mechanism. This is an ideal quantum simulator of the Heisenberg
  antiferromagnet, exposes the competition between covalent and ionic
  bonding, and can create, manipulate, and detect quantum
  entanglement. Three trapped atoms form a faithful simulator of the double
  exchange mechanism that is the fundamental building block behind many
  Heisenberg ferromagnets.  \end{abstract}


\maketitle

Recent experimental advances allow investigators to confine up to ten atoms
in a trap and address their quantum state~\cite{Cheinet08,Serwane11}. This
innovation enables the Heidelberg group to isolate two distinguishable
fermions in a one-dimensional well and tune the interaction strength to
induce fermionization~\cite{Zurn12}, presenting a unique opportunity to
study the fundamental physics of short range
repulsion~\cite{Busch98,Mora04,Idziaszek06,Liu10,Rubeni12,Rontani12,Gharashi12,Brouzos12}.
This could allow experimentalists to realize an analog to the Stoner
model for itinerant ferromagnetism~\cite{Stoner38,Bugnion13ii}. However, many
real-life solids are best described by spins that are localized in
real-space. These are commonly described by the Heisenberg model that
predicts the direct and double exchange mechanisms behind the emergence of
antiferromagnetism and ferromagnetism respectively. Here we take advantage
of the ability of the Heidelberg group to trap either two or three fermions
in the double well potential to realize the first exact quantum simulators
for the direct exchange and double exchange mechanisms. This allows us to
not only study the minimal building block of Heisenberg magnetism, but
moreover build a quantum simulator to expose the competition between
covalent and ionic bonding, and study quantum entanglement manipulation and
detection.

The direct exchange mechanism has previously been realized in double quantum
dot systems~\cite{Burkard99,Petta05,Laird06,Zhang09}, and cold atom gases in
an array of double well potentials~\cite{Trotzky08,Raizen09,Greif12}. We now
exploit the experimental flexibility of the Heidelberg group to isolate just
two atoms and expose the phase behavior. We use exact diagonalization to
deliver the full energy spectrum, and a perturbative approach to gain an
intuitive understanding of the phase behavior. The flexibility of the setup
allows us to build the first quantum model of the fundamental covalent and
ionic bonding mechanisms in molecules and crystals, allowing us to address
the long-standing question of their relative contributions to chemical
bonds~\cite{Barrow57}.  A thorough understanding of the exchange energy also
allows us to define a new protocol to create, control, and detect quantum
entanglement~\cite{Hayes07,Foot11}.

The isolated double exchange mechanism has not yet been realized
experimentally. A faithful quantum simulator of the double exchange
mechanism is not only important for understanding Heisenberg ferromagnets
but also describes the $90^{\circ}$ superexchange mechanism, and can be
extended to larger systems through a quantum virial or cluster
expansion~\cite{Liu09,Nascimbene10}.  Through exact diagonalization and a
complementary perturbative analysis we demonstrate how the cold atom gas
will probe the double exchange mechanism with changing barrier height,
interaction strength, and the ellipticity of the external trapping
potential. This modifies the degeneracy of the ground state leading to a
quantum phase transition that we expose with a statistical tunneling probe.

\section{Formalism}

We start from two-component Fermionic atoms with each species indexed by a
pseudospin $\sigma\in\{\uparrow,\downarrow\}$. The atoms are trapped by an
external harmonic potential with Hamiltonian $\hat{H}=-\nabla^2/2
+[\omega_{\perp}^2(x^2+y^2)+\omega_{\parallel}^2z^{2}]/2
+V_\mathrm{B}\exp(-\omega_{\text{B}}z^{2})
+g(\vec{r}_{\uparrow}-\vec{r}_{\downarrow})
\hat{n}_{\uparrow}(\vec{r}_{\uparrow})
\hat{n}_{\downarrow}(\vec{r}_{\downarrow})$, setting $\hbar=m=1$
throughout. The parabolic trapping potential has a variable ellipticity that
we associate with a length scale
$a_{\parallel}=1/\sqrt{\omega_{\parallel}}$. The Gaussian barrier
$V_\mathrm{B}\exp(-\omega_{\text{B}}z^{2})$ will split the system into two wells.
Throughout this paper we set its width with
$\omega_{\text{B}}=5\omega_{\parallel}$. We use two complementary techniques
to analyze the system: firstly exact diagonalization to expose the full
energy spectrum, and secondly perturbation theory to deliver an intuitive
description.

In exact diagonalization~\cite{Szabo89}, we work in the eigenbasis of the
non-interacting Hamiltonian. We express the orbitals in this basis as linear
combinations of the Gaussian-type orbitals (GTOs)~\cite{Helgaker00} of the
harmonic trapping potential without the central barrier. Labeling these GTOs
by the standard quantum numbers
$\{n_{\text{x}},n_{\text{y}},n_{\text{z}}\}$, we include
$0\le(n_{\text{x}},n_{\text{y}})<4$ and $0\le n_{\text{z}}<50$.  Including
basis functions with higher $n_\mathrm{z}$ is necessary to capture the
effect of the central barrier on the orbitals.

To calculate the orbitals of the non-interacting system including the
central barrier we first evaluate the non-interacting Hamiltonian matrix in
the GTO basis. The matrix elements of the central Gaussian barrier can be
conveniently expressed in terms of hyper-geometric
functions~\cite{Grad07}. Having calculated the Hamiltonian matrix, we
diagonalize it to obtain a list of orbitals to use in the subsequent
calculation that incorporates the effect of interactions.

We construct the $10,\!000$ Slater determinants with lowest non-interacting
energy to use as the many-body basis for our exact diagonalization
calculation.  To include interactions we evaluate the four-center integrals
in the basis of our orbitals and construct the Hamiltonian matrix using the
Slater-Condon rules~\cite{Szabo89}. Finally, we diagonalize the matrix to
find the ground and excited states of the system.

In a cold atom gas the Feshbach resonance is used to tune a contact
interaction strength $V(\vec{r})$ from the repulsive through to the
attractive regime. To generate a positive scattering length for a potential
with a small effective range we use the square well potential
$g(\vec{r})=-g\theta(R-|\vec{r}|)$ with a radius $R=0.4a_{\parallel}$. This
is significantly less than the width of the central barrier
$2/\sqrt{\omega_{\text{B}}}\approx0.9a_{\parallel}$ so that atoms localized
in adjacent wells will not interact, and moreover we ensured that the energy
of the states converged in the limit $R\to0$.  The well depth $g$ was tuned
to generate a positive or negative scattering length
$a=R[1-\tan(R\sqrt{g})/R\sqrt{g}]$, and was constrained to $0\le
R\sqrt{g}<4.49$ to confine at most one bound state. However, the inclusion
of the bound state leads to many molecular states crossing the lowest energy
open channel (see \figref{fig:DirectPlot}(a)).  This requires us to
adiabatically track states between calculations performed at neighboring
scattering lengths $a_{i}$, $a_{i+1}$, that can be uniquely followed by
connecting states with the correct spin, inversion symmetry, angular
momentum $L_{\text{z}}$, and the greatest wave function overlap
$\langle\psi_{m}(a_{i})|\psi_{n}(a_{i+1})\rangle$ where $m,n$ are state
indices.

\section{Direct exchange}

\begin{figure}
\begin{tabular}{ll} (a) Energy spectrum&(b) Tilted trap\\[-1pt]
\includegraphics[width=0.5\linewidth]{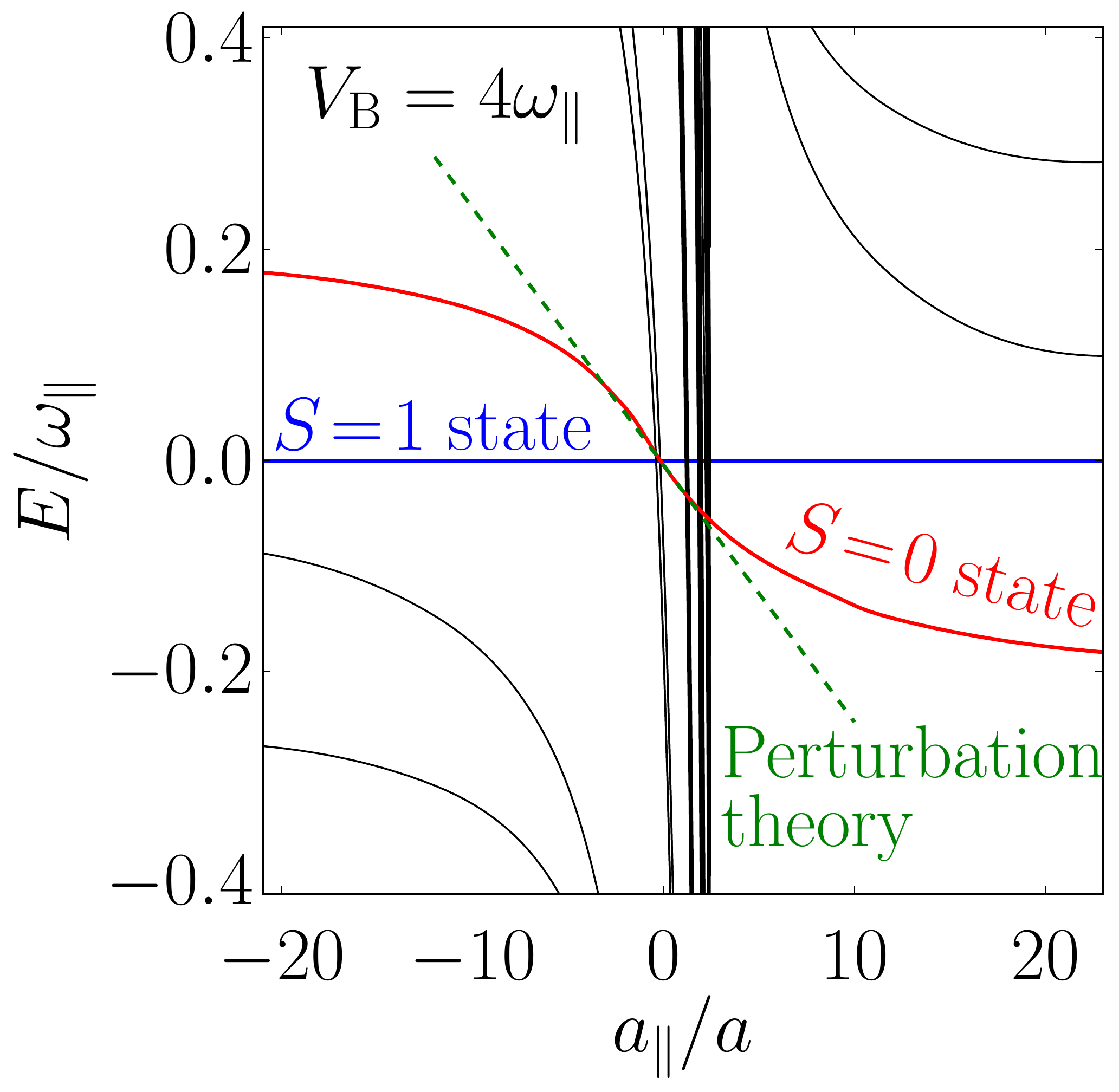}&
\includegraphics[width=0.5\linewidth]{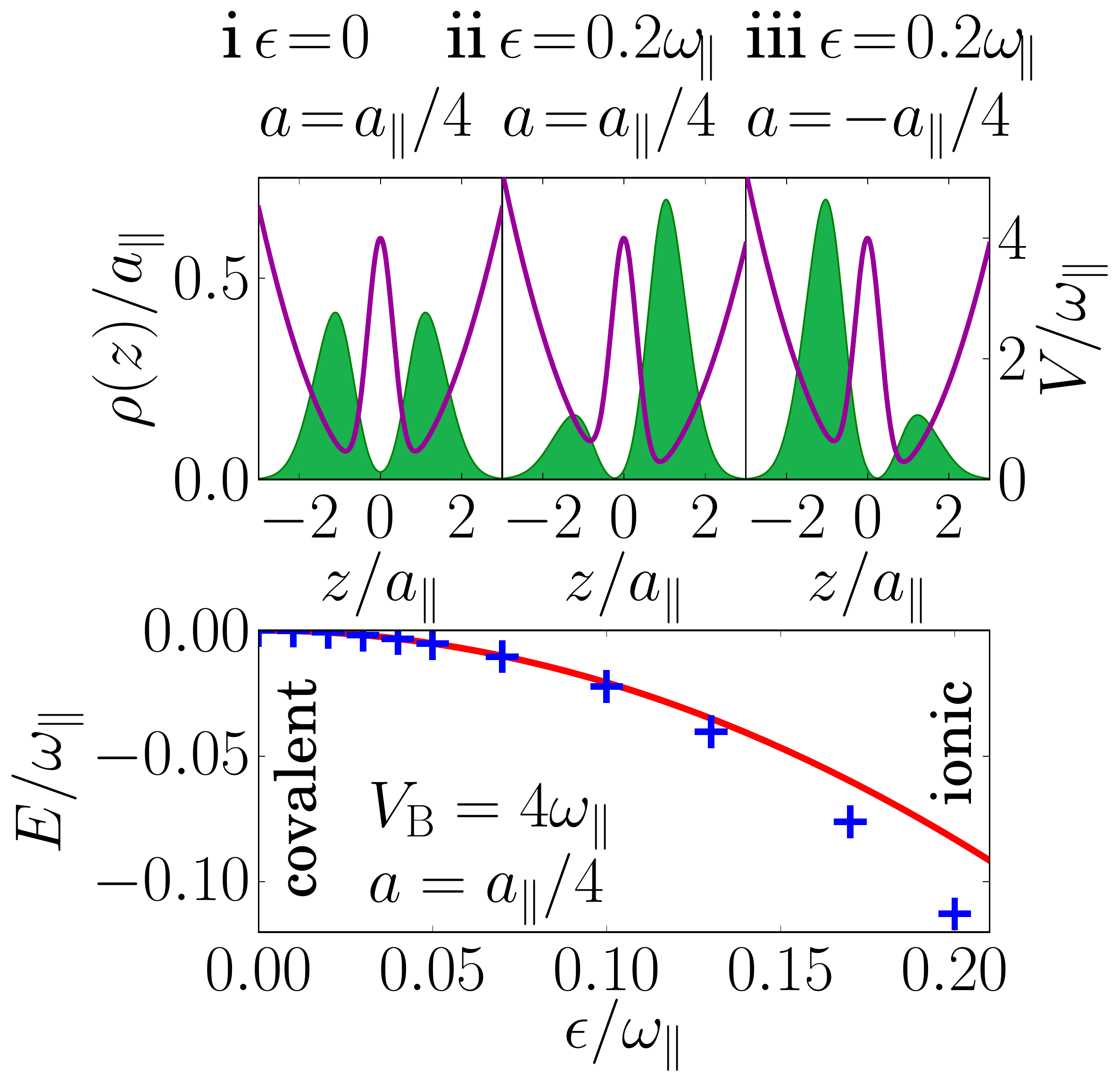}
\end{tabular}\vspace{-8pt}
\begin{flushleft}(c) Qubit control protocol\end{flushleft}\vspace{-4pt}
\includegraphics[width=1.0\linewidth]{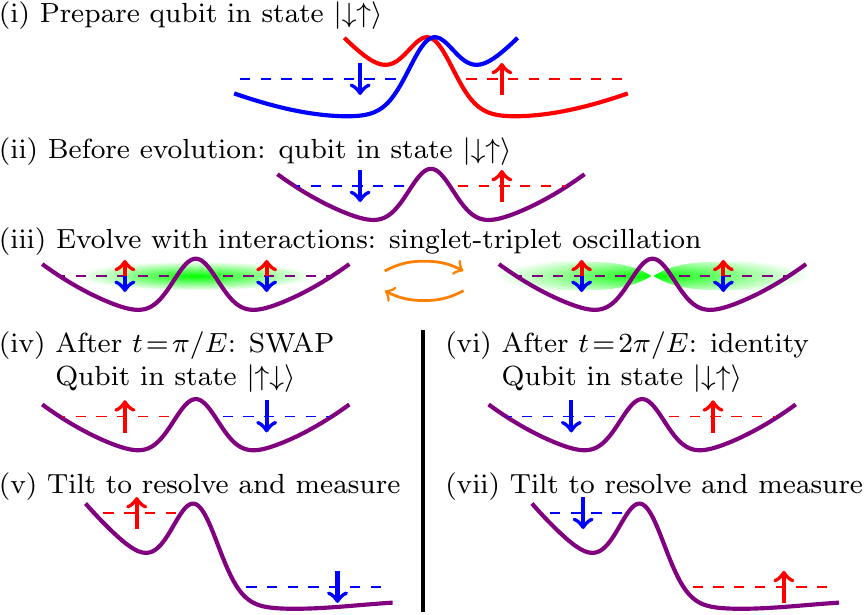}
\caption{(Color online) (a) Exchange energy from exact diagonalization
  (black lines), with the open channel in red, lowest $S=1$ state in blue,
  and perturbation theory (green dashed line).  (b) {\it Lower}:
  Normalized exchange energy of the $S=0$ level as a function of relative
  well depth $\epsilon$
  from exact diagonalization (blue crosses) and from the Hubbard Hamiltonian
  (red line). 
  {\it Upper}: The density profile (green) in the trap (magenta) at three
  values of the tilt and interaction strength.
  (c) Schematic of the experimental protocol to study quantum
  entanglement.}
\label{fig:DirectPlot} \end{figure}

We first study two fermionic atoms trapped within the double well potential.
This is a minimal quantum simulator of the direct exchange mechanism,
exposes the competition between covalent and ionic bonding, and allows us to
probe quantum entanglement. These three applications demand a thorough
understanding of the exchange energy that we study with two complementary
approaches: exact diagonalization and perturbation theory. The exchange
energy can be probed directly in experiment by applying a magnetic field
gradient and measuring the tunneling rate of atoms from the
system~\cite{Serwane11,Rontani12}. The general interacting system with no
central barrier has been studied by Busch~\cite{Busch98} so we first address
a perturbative barrier before focusing on a high barrier.

\emph{Shallow barrier}: We analyze the system analytically with first order
perturbation theory. For the $S=0$ state we start with two opposite spin
atoms in the Gaussian non-interacting ground state of the trapping potential
with no central barrier $\phi_{\sigma}(\vec{r})$, and for $S=1$ with two
equal spin atoms with one excited into a state with a node either in the
longitudinal or transverse directions depending on whether
$\omega_{\parallel}$ or $\omega_{\perp}$ is smaller. Next we introduce the
central barrier that within first order perturbation theory induces a change
in the energy of $\sum_{\sigma}\int\diffd
\vec{r}|\phi_{\sigma}(\vec{r})|^{2}V_{\text{B}}\exp(-\omega_{\text{B}}z^2)$,
and scattering that according to mean-field theory increases the energy by
$4\pi a
\int\diffd\vec{r}|\phi_{\uparrow}(\vec{r})|^2|\phi_{\downarrow}(\vec{r})|^2$. This
yields an exchange energy (energy difference between the $S=0$ and $S=1$
open channel states) of $-\Delta+oa/a_{\parallel}$ with
$\Delta=\min(\omega_{\parallel},\omega_{\perp})
-V_\mathrm{B}\{[\omega_{\parallel}/(\omega_{\parallel}+\omega_{\text{B}})]^{1/2}
-[\omega_{\parallel}/(\omega_{\parallel}+\omega_{\text{B}})]^{3/2}\}$, and
overlap matrix element $o=\sqrt{2/\pi}\omega_{\perp}
[1-4\sqrt{2}V_\mathrm{B}/\pi(\omega_{\parallel}+\omega_{\text{B}})]$. Increasing
both the barrier height and interaction strength lowers the exchange
energy. However, exact diagonalization confirms that the exchange energy is
negative for any positive scattering length, so that a negative scattering
length is required for a $S=1$ ferromagnetic ground state in conformance
with the Lieb-Mattis theorem~\cite{Lieb62}.

\emph{High barrier}: We now focus on a high barrier that presents a unique
opportunity to study a minimal Hamiltonian of localized spins. We initially
exploit exact diagonalization to deliver the exact ground and excited states
before developing an analytic theory to gain a heuristic understanding of
the system.  Finally, we demonstrate the versatility of the setup to study
the competition between covalent and ionic bonding, and quantum
entanglement.

Results for a trap with $\omega_\parallel=\omega_\perp$ and barrier with
$V_{\text{B}}=4\omega_\parallel$ are shown in
\figref{fig:DirectPlot}(a). The exchange energy flips sign across unitarity,
so that the red $S=0$ level is the ground state for $a>0$, and the blue
$S=1$ level for $a<0$. On approaching unitarity from positive scattering
lengths many molecular bands anti-cross the open channel from $0\le
a_{\parallel}/a\lesssim3$, removing the energy gap to the other
states. Although the interaction potential harbors just one bound state,
these bands correspond to the molecule being excited within the
external trapping potential. Raising the central barrier increases the
energy gap, presenting a larger range of the open channel to
experiments. This should facilitate experiments that adiabatically transit
across unitarity.  Although the two atom system cannot fall into a bound
state without violating energy conservation, the closure of the energy gap
precludes using this region to explore quantum entanglement. However, the
Super-Tonks regime is free from the molecular bands, making it the ideal
venue to use the system as a quantum simulator and study quantum
entanglement.

We next develop a complementary analytical expressions for the ground state
energy to gain an intuitive understanding of the system's behavior. We model
the system by the Hubbard Hamiltonian
$t\sum_{\sigma}(c_{\text{L}\sigma}^{\dagger}c_{\text{R}\sigma}
+c_{\text{R}\sigma}^{\dagger}c_{\text{L}\sigma})
+\epsilon\sum_{\sigma}(c_{\text{L}\sigma}^{\dagger}c_{\text{L}\sigma}
-c_{\text{R}\sigma}^{\dagger}c_{\text{R}\sigma})
+g(\ud{\text{L}}\dd{\text{L}}\cd{\text{L}}\cu{\text{L}}
+\ud{\text{R}}\dd{\text{R}}\cd{\text{R}}\cu{\text{R}})$, where
$\ud{\text{L}}$ creates a particle localized in the left-hand well,
$\ud{\text{R}}$ in the right-hand well and $\epsilon$ denotes the relative
depth of the wells.  Diagonalization in the high-barrier limit
$t\ll(\epsilon,go)$ yields an exchange energy
$-(1+\epsilon^2a_{\parallel}^{2}/16a^2o^2)t^2a_{\parallel}/4ao$, where we
use WKB perturbation theory~\cite{Platt08} to derive the parameters
$t=32\omega_{\parallel}\exp(-2\sqrt{\pi
  V_\mathrm{B}/\omega_{\text{B}}})/\pi^{2}$, and
$o=3\omega_{\perp}\sqrt{\omega_{\parallel}}/(2\pi)^{3/2}$.  We apply a
magnetic field gradient $\diffd B/\diffd z$ adding a term $\Delta H=\alpha
z$ to the Hamiltonian, where $\alpha=g_{\text{J}}\mu_{\text{B}}\diffd
B/\diffd z$ and $g_{\text{J}}$ is the $g$-factor. This gives a relative well
depth of $\epsilon\simeq[\log(2\omega_{\text{B}}
V_{\text{B}}/\omega_\parallel^2)/\omega_{\text{B}}]^{1/2}\alpha$.  We find
that the perturbative analysis matches well to the results of exact
diagonalization around unitarity, as shown in \figref{fig:DirectPlot}(a,b).

\emph{Covalent versus ionic bonding}: The covalent bonding in a crystal is
driven by the negative exchange energy of electrons being shared between
neighboring atoms. Conversely, in ionic bonding the constituent ions carry
different electronegativities, driving a displacement of electron density
that results in bond polarization and concomitant bonding.  Cold atoms present
an ideal simulator of the chemical bonding mechanisms, where the external
potential portrays the atomic pseudopotential, the fermionic atoms represent
the valence electrons, and the tilt $\epsilon$ the relative
electronegativity tuning from $\epsilon=0$ (covalent) to
$|\epsilon|\sim\omega_{\parallel}$ (ionic character).
\figref{fig:DirectPlot}(b) shows that the exchange energy is
$-(1+\epsilon^2a_{\parallel}^{2}/16a^2o^2)t^2a_{\parallel}/4ao$, where in
the absence of the tilt the first term $-t^{2}a_{\parallel}/4ao$ corresponds
to the covalent bonding, and the second term further lowers the energy in
the presence of the tilt so corresponds to the ionic contribution. On
introducing the tilt \figref{fig:DirectPlot}(b:i-ii) demonstrates how the
bond becomes polarized, with the increasing density in the lower well $\sim
a_{\parallel}^{3}t^{2}\epsilon/32a^3o^3$ lowering the net energy
quadratically $\sim\epsilon^{2}$ conforming with Pauling's definition of
electronegativity~\cite{Pauling45,Barrow57}. The lowering of the energy and
increasing polarization of the bond could be detected by tunneling atoms out
of the two sides of the trap. Finally, \figref{fig:DirectPlot}(b:iii) shows
that with $a<0$ the atomic density would be counter-intuitively pulled into
the higher well.

\emph{Quantum entanglement}: As the experimental setup offers a unique level
of experimental control we are well positioned to explore quantum
entanglement. The qubit control protocol is shown in
\figref{fig:DirectPlot}(c). We define the first qubit state as the up-spin
localized in the left-hand well and the down-spin in the right-hand well,
denoted $\ket{\downarrow\uparrow}$, \figref{fig:DirectPlot}(c:ii \& c:vi),
and the second qubit state conversely, $\ket{\uparrow\downarrow}$,
\figref{fig:DirectPlot}(c:vi).  In \figref{fig:DirectPlot}(c:i) the system
is prepared into a qubit basis state in the non-interacting regime in the
presence of a magnetic field gradient $g_{\text{J}}\mu_{\text{B}}\diffd
B/\diffd z>t/a_{\parallel}$ in the Zeeman regime so that the two spins are
driven in opposite directions~\cite{Serwane11}. With the configuration in
\figref{fig:DirectPlot}(c:i) we form the qubit state
$\ket{\downarrow\uparrow}$ in \figref{fig:DirectPlot}(c:ii). When the tilt
is removed the qubit state $\ket{\downarrow\uparrow}$ is a superposition of
the singlet and triplet states.  With interactions the system will evolve
under Rabi oscillations between the singlet and triplet states in
\figref{fig:DirectPlot}(c:iii) with a Rabi period $2\pi/E$, with
$E=-t^2a_{\parallel}/4ao$. A duration of $\pi/E$ corresponds to a SWAP
rotation~\cite{Nielsen00} into the qubit state $\ket{\uparrow\downarrow}$
shown in \figref{fig:DirectPlot}(c:iv), and a duration $2\pi/E$ an identity
rotation into the qubit state $\ket{\downarrow\uparrow}$ shown in
\figref{fig:DirectPlot}(c:vi).  To detect the qubit states in
\figref{fig:DirectPlot}(c:v) and \figref{fig:DirectPlot}(c:vii) we apply a
strong magnetic field gradient in the Paschen-Back regime to empty the
right-hand well but leave the left well occupied since it is shielded by the
central barrier. The escaped atom carries a definite spin, resolving the
system onto the qubit basis, and the remaining atomic spin can then be
measured separately following the protocol in Ref.~\cite{Serwane11} to
identify the qubit state.  Finally, we note that $\diffd
E/\diffd\epsilon|_{\epsilon=0}=0$ and therefore the Rabi period is
insensitive to the controlling tilt parameter~\cite{Stopa08}, making this an
ideal opportunity to explore entanglement in a flexible, clean, and stable
environment.

\section{Double exchange}

\begin{figure}\includegraphics[width=1.\linewidth]{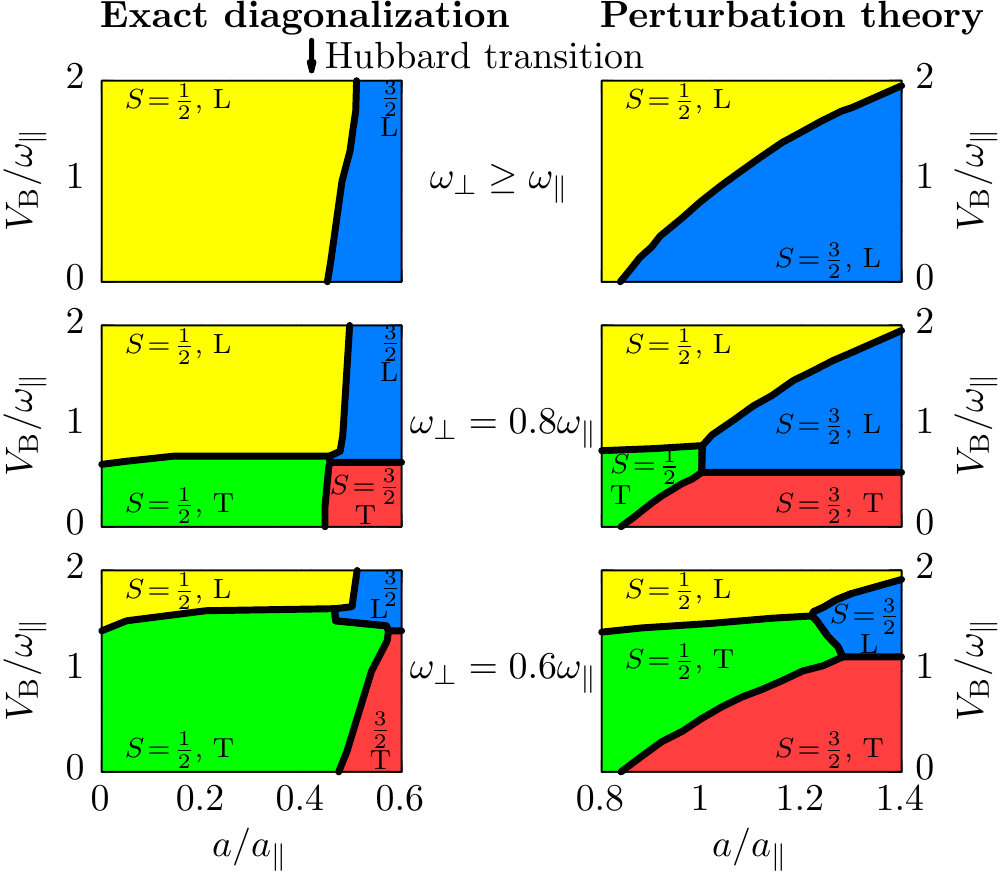}
\caption{(Color
online) The phase diagrams predicted by exact diagonalization (left) and
perturbation theory (right) at three different trap ratios. The phase corresponding to
the each color is identified in each diagram, with ``L''
indicating a longitudinal phase and ``T'' a transverse phase. The arrows
show the critical interaction strength predicted by the Hubbard model.}
\label{fig:DoublePlot}
\end{figure}

\begin{table*} \begin{tabular}{l|l|c|c|l}
Spin&State(s)&Orientation&Degeneracy&Energy\\\hline\hline
\multirow{3}{*}{$S=\frac{1}{2}$}& $\dd{000}\ud{000}\ud{001}\vac$& Longitudinal&
1& $\frac{5}{2}\omega_{\parallel}+3\omega_{\perp}
+\sqrt{\frac{\omega_{\parallel}}{\omega_{\parallel}+\omega_{\text{B}}}}V_\mathrm{B}\left(2
+\frac{\omega_{\parallel}}{\omega_{\parallel}+\omega_{\text{B}}}\right)
+\frac{a}{a_{\parallel}}\omega_{\perp}\sqrt{\frac{2}{\pi}}
\left(\frac{3}{2}-\frac{4\sqrt{2}}{\pi}\frac{V_\mathrm{B}}{\omega_{\parallel}+\omega_{\text{B}}}\right)$\\\cline{2-5}
&$\dd{000}\ud{000}\ud{010}\vac$& \multirow{2}{*}{Transverse}&
\multirow{2}{*}{2}&
\multirow{2}{*}{$\frac{3}{2}\omega_{\parallel}+4\omega_{\perp}
+3\sqrt{\frac{\omega_{\parallel}}{\omega_{\parallel}+\omega_{\text{B}}}}V_\mathrm{B}
+\frac{a}{a_{\parallel}}\omega_{\perp}\sqrt{\frac{2}{\pi}}
\left(\frac{3}{2}-\frac{6\sqrt{2}}{\pi}\frac{V_\mathrm{B}}{\omega_{\parallel}+\omega_{\text{B}}}\right)$}\\
&$\dd{000}\ud{000}\ud{100}\vac$&&&\\\hline \multirow{3}{*}{$S=\frac{3}{2}$}&
$\ud{000}\ud{001}\ud{010}\vac$& \multirow{2}{*}{Longitudinal}&
\multirow{2}{*}{2}&
\multirow{2}{*}{$\frac{5}{2}\omega_{\parallel}+4\omega_{\perp}
+\sqrt{\frac{\omega_{\parallel}}{\omega_{\parallel}+\omega_{\text{B}}}}V_\mathrm{B}\left(2
+\frac{\omega_{\parallel}}{\omega_{\parallel}+\omega_{\text{B}}}\right)$}\\
&$\ud{000}\ud{001}\ud{100}\vac$&&&\\\cline{2-5}
&$\ud{000}\ud{010}\ud{100}\vac$& Transverse& 1&
$\frac{3}{2}\omega_{\parallel}+5\omega_{\perp}
+3\sqrt{\frac{\omega_{\parallel}}{\omega_{\parallel}+\omega_{\text{B}}}}V_\mathrm{B}$\\\hline
\end{tabular} \caption{Lowest energy states for the three atom system and
  their energy calculated in perturbation theory. Both the $S=1/2$ and $S=3/2$
  states are shown, and also the available longitudinal and transverse
  modes.} \label{tab:WaveFunctions} \end{table*}

Two trapped atoms with a positive scattering length always yields a $S=0$
ground state \cite{Lieb62}, so to realize the simplest possible
ferromagnetic ground state we turn to a three-atom system. For a
sufficiently strong central barrier this delivers a quantum simulator of the
double exchange mechanism for ferromagnetism: two atoms are localized in the
lowest level of each well, and a third itinerant atom in a higher energy
level couples their spins. With three atoms there are two possibilities for
the highest occupied orbital: either with a node longitudinally
($n_{\text{z}}=1$), or two degenerate modes with a transverse node
($n_{\text{x},\text{y}}=1$).  To study the emergence of magnetic
correlations we use both exact diagonalization and perturbation
theory. We adopt the same perturbation theory as introduced in the direct
exchange section for a small central barrier and weak scattering, except now
for the three-atom states shown in \tabref{tab:WaveFunctions}. To orient the
discussion we focus initially on the realization of itinerant ferromagnetism
before introducing a central barrier to consider the double exchange
mechanism.

Exact diagonalization delivers the phase diagrams in \figref{fig:DoublePlot}
that compare well with the perturbation theory predictions. At
$V_\mathrm{B}=0$ the system enters the $S=3/2$ state at
$a/a_{\parallel}\approx0.45$, which compares favorably to the perturbation
calculation estimate that $a/a_{\parallel}=\sqrt{2/\pi}/3\approx0.84$ and a
prediction that we extrapolated from the data of Liu of
$a/a_{\parallel}\approx1.1$~\cite{Liu10}. We note that the exact theory has
a lower critical interaction strength than the perturbation theory
prediction, which reflects the situation of the itinerant
case~\cite{Conduit08,Conduit09}. The critical interaction strength is
predicted to be independent of trap ellipticity
$\omega_{\perp}/\omega_{\parallel}$ by both exact diagonalization and the
perturbation theory calculations. In the high barrier potential limit we
follow the example from the direct exchange calculation and model the system
with the Hubbard Hamiltonian, predicting a crossover at
$a/a_{\parallel}=\sqrt{\pi/2}/3\approx0.42$. This is in direct agreement
with the results from exact diagonalization.

When $\omega_{\parallel}<\omega_{\perp}$ it is energetically favorable to
occupy longitudinal modes and when $\omega_{\parallel}>\omega_{\perp}$ the
transverse modes are preferred. However, the introduction of a central
barrier favors longitudinal states ($n_{\text{z}}=1$) with a node
across the barrier, with the crossover at
$V_{\text{B}}=\sqrt{1+\omega_{\text{B}}/\omega_{\parallel}}
(1+\omega_{\parallel}/\omega_{\text{B}})(\omega_{\parallel}-\omega_{\perp})$
for both $S=1/2$ (at $a=0$) and $S=3/2$. A weak central barrier forces the
atoms apart reducing the effective interaction strength, meaning that the
boundary between the $S=1/2$ and $S=3/2$ phases has a positive slope of
$\sqrt{\omega_{\parallel}}(\omega_{\parallel}+\omega_{\perp})/24\sqrt{\pi}$.
With a larger central barrier the longitudinal $S=1/2$ state and transverse
$S=3/2$ state share a phase boundary with a negative slope giving rise to a
characteristic notch.

The phases are distinguished not only by their spin quantum number but also
the degeneracy associated with the orientation of the node in the highest
occupied orbital. The changing degeneracy means that the phases with the
same spin quantum number cannot evolve into one another so there is a
quantum phase transition between the two. In a cold atom gas both total spin
and symmetry are conserved, so to probe the phase diagram a tunneling
measurement is proposed~\cite{Serwane11}, with the gas starting from four
atoms, tilting the trap, and tunneling down to the three atom configuration
with the lowest energy. At the critical interaction strength the $S=1/2$ and
$S=3/2$ states will be formed with equal likelihood, but taking account of
the degeneracy will boost the formation of the doubly degenerate states,
presenting an ideal tool to characterize the phases.

The high barrier perturbation theory provided a perfect description for the
phase behavior predicted by exact diagonalization theory in that limit. This
demonstrates that the ferromagnetism is driven by double exchange,
consistent with the observed exponential decay of the interaction matrix
element between one atom localized in the left and another in the right-hand
well. However, the high barrier perturbation theory is unable to describe
the system with a low central barrier, where the requirement to have
orbitals with a longitudinal node is incompatible with the node-less
``transverse'' wave functions found in exact diagonalization and shown in
\tabref{tab:WaveFunctions}, which are of itinerant nature.  Therefore, the
crossover from ``transverse'' to ``longitudinal'' wavefunctions in the phase
diagram denotes the underlying ferromagnetic correlations changing from
itinerant to localized double exchange.

\section{Discussion}

The double well system is the perfect playground to study the exchange
mechanisms of strongly interacting fermions. We have studied the direct
exchange mechanism that is the fundamental building block of Heisenberg
antiferromagnetism, and exposes the competition between covalent and ionic
bonding. The experimental flexibility of the cold atom gas also presents the
ideal arena to study quantum entanglement. Trapping three fermions delivers
the first faithful realization of the double exchange mechanism. This can
provide insights into Heisenberg ferromagnets and $90^{\circ}$
superexchange, and be built up to larger lattices through a cluster
expansion.

The exchange mechanisms presented here give a tantalizing insight into the
broad range of effects that can be explored in the double well
potential. The inclusion of a third potential well would allow investigators
to study the Kramers-Anderson superexchange mechanism~\cite{Anderson50}
behind many antiferromagnets. Trapping more atoms should reveal an even/odd
effect of flipping between antiferromagnetic and ferromagnetic ground
states.

\acknowledgments{The authors thank Gerhard Z\"urn, Thomas Lompe, Selim
  Jochim \& Stefan Baur for useful discussions. POB acknowledges the
  financial support of the EPSRC, and GJC the support of Gonville \& Caius
  College.}

\end{document}